\documentstyle[aps,twocolumn,prl,epsfig]{revtex}

\newcommand{\be}{\begin{equation}}
\newcommand{\ee}{\end{equation}}
\newcommand{\br}{\begin{eqnarray}}
\newcommand{\er}{\end{eqnarray}}
\newcommand{\ba}{\begin{array}}
\newcommand{\ea}{\end{array}}
\newcommand{\bi}{\begin{itemize}}
\newcommand{\ei}{\end{itemize}}
\newcommand{\bn}{\begin{enumerate}}
\newcommand{\en}{\end{enumerate}}
\newcommand{\bc}{\begin{center}}
\newcommand{\ec}{\end{center}}

\newcommand{\gsim}{\lower.7ex\hbox{$\;\stackrel{\textstyle>}{\sim}\;$}}
\newcommand{\lsim}{\lower.7ex\hbox{$\;\stackrel{\textstyle<}{\sim}\;$}}

\def\sq{\tilde{q}}

\def\ar{\to}

\begin{document}
\twocolumn[\hsize\textwidth\columnwidth\hsize\csname @twocolumnfalse\endcsname


\title{ New faces of Supersymmetry:
effects of large phases on Higgs production}
\author{A. Dedes$^1$ and  S. Moretti$^{1,2}$}
\address{ $^1$Rutherford Appleton Laboratory
Chilton, Didcot, Oxon OX11 0QX, UK}
\address{ $^2$Department of Radiation Sciences,
Uppsala~University, P.O.~Box~535,~75121~Uppsala,~Sweden} 

\maketitle
\begin{abstract}
If the soft Supersymmetry (SUSY) breaking masses and couplings are complex
and cancellations do take place in the SUSY induced contributions to the 
fermionic Electric Dipole Moments (EDMs), then the CP-violating soft 
phases can drastically modify much of the known phenomenological pattern of 
the Minimal Supersymmetric Standard Model (MSSM). In particular, the 
squark loop content of the dominant Higgs production mechanism at the Large
Hadron Collider (LHC), the gluon-gluon fusion mode, could be responsible
of large corrections to the known cross sections.
\end{abstract}

\vspace*{3mm}
]

\noindent
The strong constraints arising from the measurements 
of the electron and neutron EDMs
on the size of the CP-violating phases associated to
the soft SUSY Lagrangian \cite{nir}
can be evaded, if the corresponding masses and couplings arrange themselves 
so that the SUSY contributions to the EDMs cancel out. 
This has been proved to occur over a 
sizable area of the MSSM parameter space \cite{NATH,rosiek}. 
Under these circumstances, one ought to consider possible phenomenological
effects  of such `explicit' CP-violation in the soft SUSY breaking
sector \cite{Kane}.

\noindent
Higgs physics is perhaps the primary interest behind the construction
of the LHC. Within the MSSM, with or without phases, the mass
of the lightest Higgs boson, $h^0$, is expected to be well
within the reach of the future CERN hadron collider. However, the
dominant production mode of this particle (and of the other two neutral Higgs
bosons of the theory, $H^0$ and $A^0$) can be affected
by a non-zero value of either of the two independent 
phases, $\phi_\mu$ and $\phi_A$,
associated to the (complex) Higgsino mass term, $\mu$, and trilinear scalar
couplings
$A\equiv A_u = A_d$, where $u$ and $d$ refer
 to all flavours of up- and down-type (s)quarks, respectively. 
In fact,
the production of one on-shell Higgs boson via gluon-gluon fusion 
\cite{HHG} proceeds
through loops of both quarks and squarks (primarily, those of top
and bottom flavour). By a close look at the 
squark-squark-Higgs vertices 
(which we collectively denote by 
$\lambda_{\Phi^0{\tilde{q}}_\chi{\tilde{q}}^*_{\chi'}}$, with
$\Phi^0=h^0,H^0,A^0$ and 
$q=u,d$ -- here,
we are only interested in vertices involving neutral Higgs bosons and 
the combination $\chi=\chi'$: see
\cite{bigone} for $\chi\ne\chi'$ and/or charged Higgs scalars)
in the chiral (or weak) basis of
Ref.~\cite{HHG}  
(i.e., $\chi,\chi'=L,R$) and at the mixing relations\footnote{These
originally appeared in the first paper of \cite{wagner}.}
converting the latter into the mass basis (i.e., $\chi,\chi'=1,2$), i.e.,
\begin{eqnarray}\label{mixing}
\lambda_{\Phi^0\tilde{q}_1\tilde{q}^{*}_1} &=& c_{\sq} c_{\sq} 
\lambda_{\Phi^0\tilde{q}_L\tilde{q}^{*}_L} + s_{\sq} s_{\sq}
\lambda_{\Phi^0\tilde{q}_R \tilde{q}^{*}_R}\nonumber \\ &+&
c_{\sq} s_{\sq} e^{i \phi_{\tilde{q}}}
\lambda_{\Phi^0\tilde{q}_L\tilde{q}^{*}_R} + s_{\sq} c_{\sq} 
e^{-i \phi_{\tilde{q}}}
\lambda_{\Phi^0\tilde{q}_R\tilde{q}^{*}_L}, \nonumber \\
\lambda_{\Phi^0\tilde{q}_2\tilde{q}^{*}_2} &=& s_{\sq} s_{\sq} 
\lambda_{\Phi^0\tilde{q}_L\tilde{q}^{*}_L} + c_{\sq} c_{\sq}
\lambda_{\Phi^0\tilde{q}_R\tilde{q}^{*}_R} \nonumber \\
&-& s_{\sq} c_{\sq} e^{i \phi_{\tilde{q}}}
\lambda_{\Phi^0\tilde{q}_L\tilde{q}^{*}_R}
-c_{\sq} s_{\sq} e^{-i \phi_{\tilde{q}}}
\lambda_{\Phi^0\tilde{q}_R\tilde{q}^{*}_L}, 
\end{eqnarray}
it is clear that $\phi_\mu$ and $\phi_A$ end up into the squark
loop contributions to $gg\to \Phi^0$, via $\phi_{{\tilde{q}}}$, 
the phases associated to the soft squark masses, in turn expressed
in terms of the previous two. (We follow the notation of 
 Ref.~\cite{bigone}.) Here, $c_{\tilde{q}}$
and $s_{\tilde{q}}$ are the cosine and sine of the mixing angle
$\theta_{\tilde{q}}$ entering the unitary transformation that 
diagonalises the squark mass matrix (alongside $\phi_{{\tilde{q}}}$).
It is the purpose of this letter to assess the extent of the
corrections induced to the total cross sections of $gg\to \Phi^0$
(for any Higgs state) at the
LHC by finite values of $\phi_\mu$ and $\phi_A$. 

\noindent
In order to do so, we proceed as follows. First, we establish which
are the combinations of MSSM parameters that guarantee the mentioned
cancellations among the SUSY contributions to the EDMs.
Then, we enforce the current collider limits on the 
squark and Higgs masses and couplings concerned: primarily, those
of the lightest Higgs scalar, $h^0$, and squark, ${\tilde{t}}_1$. 
(Some points will also be excluded
by the requirement of positive definiteness of the squared squark masses.) 
Finally, we compute the $gg\to\Phi^0$ rates with and without phases
and plot the ratio between the two results. We do so at 
leading order (LO) and include only the top and bottom (i.e.,
$t$ and $b$) and stop and sbottom (i.e., $t_1,t_2$ and $b_1,b_2$, with $1,2$
in order of increasing mass)
loops, indeed the dominant terms \cite{DDS}. At this accuracy, such a ratio 
coincides with that taken between the matrix elements themselves, as the
dependence upon the gluon distribution functions cancels out (further assuming
that the relevant hard scale is the same in both cases, e.g., 
$Q\equiv M_{\Phi^0}$). We are of course aware 
that higher order QCD corrections to the gluon-gluon fusion mode 
are very large in the MSSM \cite{DDS}. However,
it has been shown that they affect the quark and squark contributions
very similarly \cite{DDS}. Thus, we leave them aside for 
the time being. ({A two-loop analysis is performed in 
Ref.~\cite{bigone}.) 

\noindent
Before proceeding to the computation though,
 a subtlety should be noted.
The production of the pseudoscalar Higgs boson, $A^0$, proceeds at LO 
via quark loops only, if $\phi_\mu =\phi_A=0$.
In fact, for a `phaseless' MSSM, one gets that 
$\lambda_{A^0\tilde{q}_1\tilde{q}^{*}_1} =
 \lambda_{A^0\tilde{q}_2\tilde{q}^{*}_2} =0$,
 as can be deduced from 
eq.~(\ref{mixing}) if one recalls that reverting
the chirality flow in the vertex $\lambda_{A^0\tilde{q}_\chi
\tilde{q}_{\chi'}}$, with $\chi\ne\chi'=L,R$,
corresponds to changing the sign in the  Feynman rule
\cite{HHG}:
$\lambda_{A^0\tilde{q}_L\tilde{q}_R}=-\lambda_{A^0\tilde{q}_R\tilde{q}_L}$.
That the above couplings are identically zero is no longer true if either  
$\phi_\mu$ or $\phi_A$ is non-zero.
Therefore, a novel effect in the case $\Phi^0=A^0$, due to the presence of
CP-violating phases, is the
very existence of squark loop contributions to the 
amplitude associated to pseudoscalar Higgs boson production:
\widetext
\begin{equation}
{\cal M}_{ab}^{A^0}\propto
\frac{\alpha_s(Q)}{2\pi}\delta_{ab} \epsilon_\mu(P_1)\epsilon_\nu(P_2)
\Biggl \{i \varepsilon^{\mu\nu\rho\sigma}P_{1\rho}P_{2\sigma}\sum_q
\frac{\lambda_{A^0q\bar q}}{m_q} \tau_q \biggl [ f(\tau_q) \biggl ] +
\sum_{\tilde{q}}\frac{\lambda_{A^0\tilde{q}\tilde{q}^*}}
{4m_{\tilde{q}}^2} \biggl (g^{\mu\nu}P_1\cdot P_2-P_1^\nu P_2^\mu\biggr )
\tau_{\tilde{q}}\biggl[1-\tau_{\tilde{q}}f(\tau_{\tilde{q}})\biggr ]\Biggr \}. 
\label{oddme}
\end{equation}
\narrowtext
\vspace{-6mm}
\noindent
Here, $P_1$, $P_2$
are the gluon four-momenta, $\epsilon_\mu(P_1)$, $\epsilon_\nu(P_2)$ their
polarisation four-vectors and $a,b$ their colours,
$\alpha_s(Q)$ is the strong coupling constant, 
$\lambda_{A^0u\bar u}=-gm_u\cot\beta/2M_W$ and 
$\lambda_{A^0d\bar d}=-gm_d\tan\beta/2M_W$ are the standard MSSM 
quark-quark-Higgs couplings (they are affected by the
presence of the phases only in higher orders \cite{wagner}),  
$g^2=e^2/\sin^2\theta_W=4\pi\alpha_{\rm{EM}}/\sin^2\theta_W$,
$\tau_{q,\tilde{q}}=4m_{q,\tilde{q}}^2/M_{A^0}^2$ with $m_{q}$ and
$m_{\tilde{q}}$ the quark and squark masses entering the loops, respectively,
whereas $f(\tau)$ can be found in \cite{DDS}.
Furthermore, there exist no interference terms between quark and squark loops
if $\Phi^0=A^0$. In fact, in eq.~(\ref{oddme}), one can recognise
 an antisymmetric part -- $\varepsilon$ is the Levi-Civita tensor,
generated by the $\gamma^5$ matrix in the quark-quark-Higgs
vertex -- associated to the former
(first term on the right-hand side) and a
symmetric one associated to the latter (second term 
on the right-hand side). In other words, in the case of pseudoscalar
Higgs boson production, 
the SUSY corrections are always {positive}. In contrast,
see Ref.~\cite{DDS}, the SUSY terms can interfere with the Standard Model
ones in scalar Higgs boson production, i.e., $\Phi^0=h^0,H^0$, 
so that finite values of $\phi_\mu$ and $\phi_A$ can either enhance or 
deplete the phaseless MSSM production rates.

\noindent
The current limits -- at 90\% confidence level (CL) --
on the electron, $d_e$ \cite{de}, and neutron, $d_n$
\cite{dn}, EDMs are:
$|d_e| \le 4.3 \times 10^{-27} \; e \, {\rm cm}$ and
$|d_n| \le 6.3 \times 10^{-26} \; e \, {\rm cm}$. 
%
Large values of $\phi_\mu$ and $\phi_A$ are consistent with these
bounds (both in the `constrained' and `unconstrained' MSSM)
provided cancellations  take place between the contributions proportional
to the former and those proportional to the latter \cite{NATH,rosiek}. 
This certainly 
requires a certain amount of `fine-tuning' among the soft SUSY
masses and couplings \cite{rosiek}. However, it has recently 
been suggested that such cancellations occur naturally 
in the context of Superstring models \cite{brhlik1}.
Here we should point out that we are working in the 
region of the parameter space where the phases of the gaugino
masses and those of the vacuum expectation values are zero.
Also, for the neutron EDM calculation we take into account the electric, 
chromoelectric and gluon-chromoelectric dipole moment contributions 
evaluated at the electroweak (EW) scale~\cite{NATH,rosiek}.
To search for
those combinations of soft sparticle masses and couplings that guarantee 
vanishing SUSY contributions to the EDMs for each possible choice of the
CP-violating phases, we scan over the  
($\phi_\mu, \phi_A$) plane 
and use the program
of Ref.~\cite{rosiek}.
This returns those minimum values of
the modulus of the common trilinear coupling, $|A|$, 
above which the cancellations work.
These can be found in 
Fig.~\ref{fig:A} in the form of a contour plot over 
the ($\phi_\mu, \phi_A$) plane.
There, we have also superimposed those regions 
(to be excluded from further consideration)
\vspace*{12mm}
\begin{figure}[htp]%
\begin{center}
\epsfig{figure=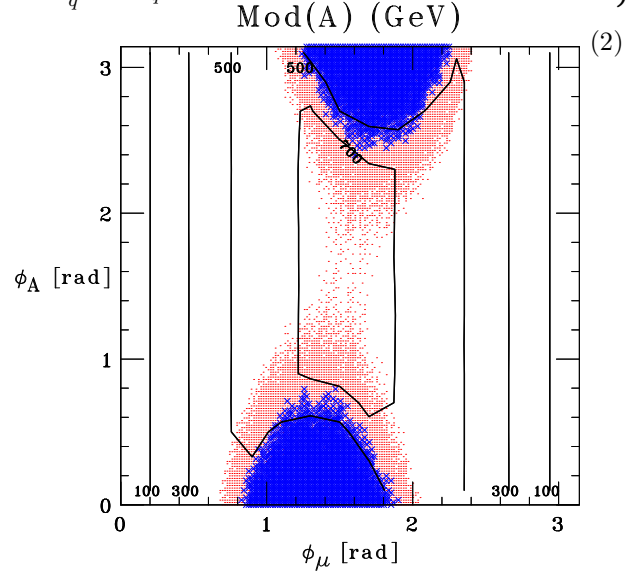,height=3.0in,angle=90}
\caption{Contour plot illustrating the minimum values of the modulus
of the common trilinear coupling, $|A|$ (consistent with
cancellations taking place in the MSSM contributions to the EDMs), for any
 combination of $\phi_\mu$ and $\phi_A$.
Dotted and crossed points indicate -- here and in the following figures --
regions which are excluded from
direct searches 
and negativity of the squared squark masses, respectively.}
\label{fig:A}
\end{center}
\end{figure}
\vspace*{-3.0mm}
\noindent
over which the observable MSSM parameters 
assume values that are either
forbidden by collider limits (dots, specifically,
on the lightest stop mass: see Fig.~\ref{fig:stop1} below)
or for which the squared squark masses
become negative (crosses), for a given 
combination of the other soft SUSY breaking parameters.
These are $|\mu|$, which is taken to be 500 GeV,
 the soft squark masses of the 
three generations $M_{\tilde{q}_{1,2,3}}$, for which
we assume -- in the notation of Ref.~\cite{HHG} --
$M_{\tilde{q}_{1,2}} \gg M_{\tilde{q}_{3}}$, 
$M_{\tilde{q}_{1,2}}$ $\equiv$ $M_{\tilde{Q}_{1,2}}$ $=$ 
$M_{\tilde{U}_{1,2}}$ $=$
$M_{\tilde{D}_{1,2}}=$~2~TeV and
$M_{\tilde{q}_3}$ $\equiv$ $M_{\tilde{Q}_3}$ $=$
$M_{\tilde{U}_3}$ $=$ $M_{\tilde{D}_3}=$ 300 GeV,
and the gluino soft mass $M_{\tilde{g}}=$ 1~TeV. 
In addition, in order to completely define our model for the calculation of 
the $gg\to \Phi^0$ processes, we also have introduced a possible choice of
the Higgs sector parameters: i.e., the  mass of one  
physical state, e.g., $M_{A^0}=200$ GeV, and
the ratio of the vacuum expectation values of the two doublet fields, e.g.,
$\tan\beta=3$. We will adopt the above numbers  as default
in the reminder of our analysis. Apart from complying with the limits
on the two-loop Barr-Zee type graphs \cite{BZ},
they should serve the sole purpose 
of being an example of the rich phenomenology that can be
induced by the CP-violating phases in the MSSM, rather than a benchmark
case. Indeed, similar effects to those illustrated below can be observed
for other choices of  $|\mu|$,  
$M_{\tilde{q}_{1,2,3}}$,  $M_{A^0}$ and $\tan\beta$~\cite{bigone}. Finally,
notice that, starting from these parameter values, 
one can verify that the heaviest squark masses, 
$m_{\tilde{t}_2}$, $m_{\tilde{b}_1}$, and $m_{\tilde{b}_2}$,
are all consistent with current experimental bounds.
As for the lightest stop, 
we display in Fig.~\ref{fig:stop1} the values attained by $m_{\tilde{t}_1}$
over the usual ($\phi_\mu,\phi_A$) plane. As a matter of fact, over most of the
latter, $m_{\tilde{t}_1}$ is well above
the current experimental reach, whose upper limit can safely be drawn 
at 120 GeV or so, given our $\tan\beta$ \cite{Tevatron}.
Also, for the above choice of $\tan\beta$ and $M_{A^0}$, one gets
that $M_{h^0}\gsim90$ GeV, in accordance with the latest
bound from LEP, of about 85.5 GeV for $\tan\beta\ge1$ at 95\% CL \cite{LEP}, 
whereas $M_{H^0}$ is approximately degenerate with $M_{A^0}$. In this respect,
notice that, since the SUSY loop corrections to the lightest Higgs boson
mass are significant, $M_{h^0}$ in general depends upon $A$ 
(see the last paper of \cite{wagner}). As such dependence
is not yet known explicitly, we have mimicked it by adopting 
 two values for $M_{h^0}$,
 within 10 GeV of the one-loop result,
 for each $|A|$ over the ($\phi_\mu,\phi_A$) plane.
In contrast, one may assume little dependence of $M_{H^0}$ upon
$A$, and thus use a unique value for it, given the negligible size of
the higher order corrections here.
%
%
%
\begin{figure}[htb]
\epsfig{figure=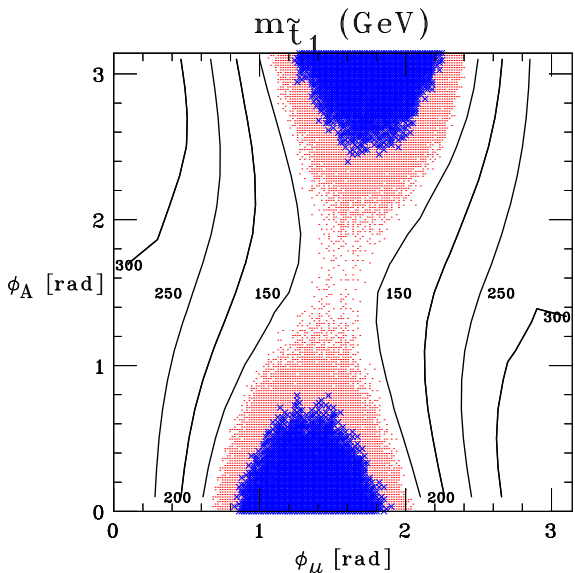,height=3.0in,angle=90}
\caption{Contour plot illustrating the values of the lightest stop mass, 
$m_{\tilde{t}_1}$,
for any combination of $\phi_\mu$ and $\phi_A$.}
\label{fig:stop1}
\end{figure}
\noindent
We now proceed to displaying the ratio:
\br
R(g g \ar \Phi^0) \ =\  
\frac{\sigma^{\mathrm{MSSM}^*}_{\mathrm{LO}}(g g \ar \Phi^0)}
     {\sigma^{\mathrm{MSSM}  }_{\mathrm{LO}}(g g \ar \Phi^0)},
\label{rat}
\er
where MSSM$^*$ refers to the case of the MSSM in presence of
CP-violating phases. (Of course, if $\phi_\mu=\phi_A=0$, 
then $R(g g \ar \Phi^0)$ is equal to 1.)

\noindent
 Fig.~\ref{fig:h0} shows the ratio in eq.~(\ref{rat})
for the case $\Phi^0=h^0$, again as a contour plot over the 
 ($\phi_\mu,\phi_A$) plane. 
One can see that
the effects of the CP-violating phases are large indeed.
\begin{figure}[p]
\epsfig{figure=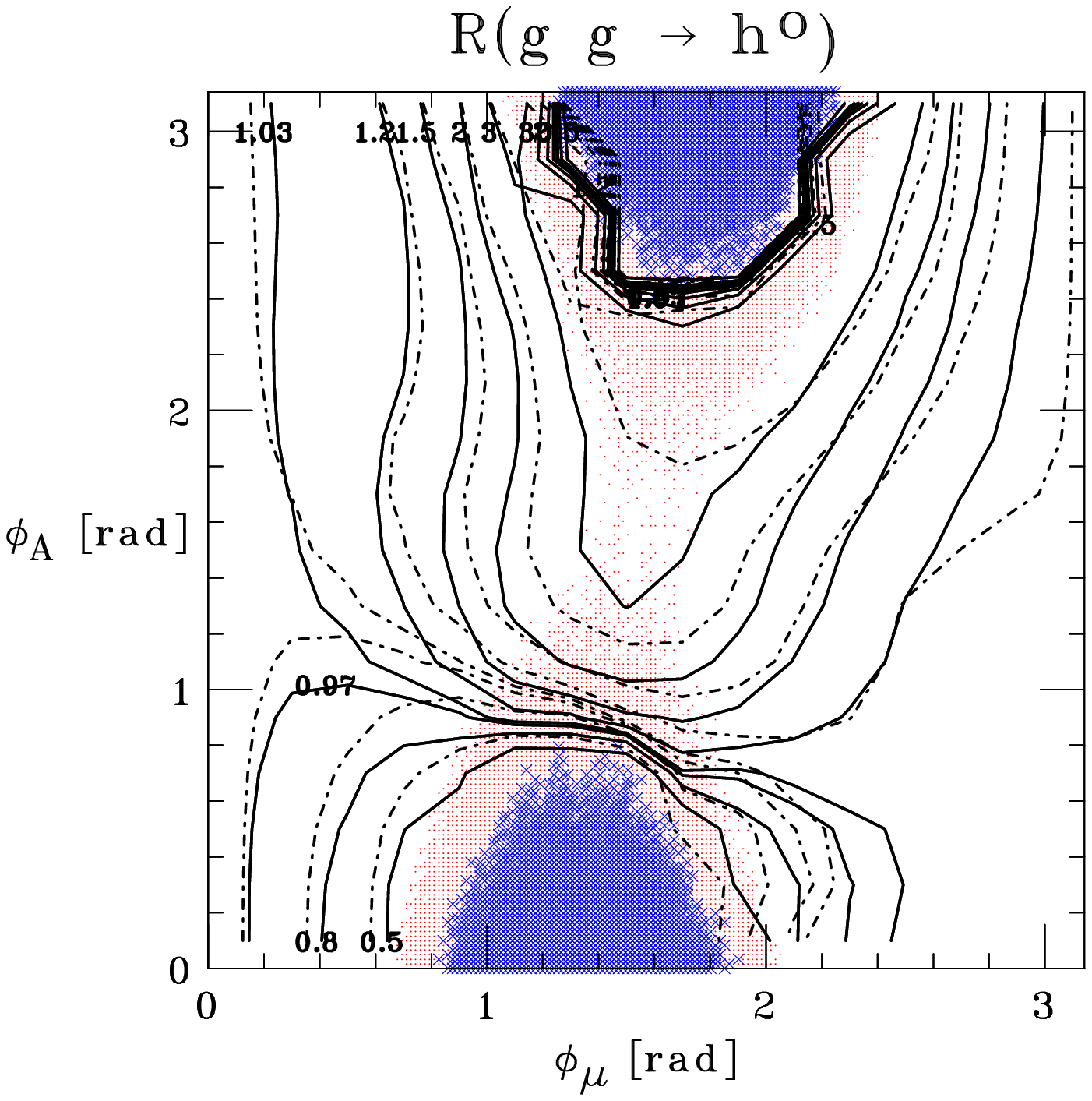,height=3.0in,angle=90}
\caption{Contour plot illustrating the values of the ratio 
in eq.~(\ref{rat}) for the case $\Phi^0=h^0$,
for any combination of $\phi_\mu$ and $\phi_A$, 
when $M_{h^0}=90$ (solid) and $100$ (dot-dashed) GeV.
}
\label{fig:h0}
\end{figure}
\noindent
Over the allowed ($\phi_\mu,\phi_A$) regions, they deplete or increase
the cross section obtained in the phaseless MSSM  by as much as
a factor of 2 and 3, respectively. 
In fact, one can distinguish two
complementary regions: $\phi_A \lsim \pi/3$ and $\phi_A \gsim \pi/3$
 (for any $\phi_\mu$).
In the first one, the effects of the phases are destructive; in the second
one, constructive. A simple explanation
for this is that $\lambda_{h^0\tilde{t}_1\tilde{t}_1^*}$ changes its
sign when $\phi_A\approx\pi/3$. 
Fig.~\ref{fig:H0} presents the rates for the case $\Phi^0=H^0$. Here too,
effects of finite values of $\phi_\mu$ and $\phi_A$ can be dramatic,
no less than in the previous case. 
The typical rates for the ratio in eq.~(\ref{rat})
are in the interval $0.5 \lsim  R(g g \ar H^0) \lsim 4 $. 
The dependence of the $H^0$ cross section on the relative value
of $\phi_\mu$ and $\phi_A$ is difficult to discern. 
In Fig.~\ref{fig:A0}, we display the pattern of eq.~(\ref{rat})
when $\Phi^0=A^0$. As already explained, one always has that 
$R(g g \ar A^0)\ge 1$. Once again, 
the ratio can become as large as 4. In this case,
a visible trend is that $R$ approaches 1 when
$\phi_A=\phi_\mu \simeq {\pi}/{2}$, as 
the coupling $\lambda_{A^0\tilde{t}_1
\tilde{t}_1^*}$ intervening in the lightest stop squark loop   
becomes zero. 
Three general remarks for all three ratios are the following.
Firstly, we have explicitly verified that
significant contributions to the total cross sections come
only from top, bottom  and lightest stop loops.
Secondly, the effects of the phases are more evident where $|A|$ is
larger, because of its intervention in the 
$\lambda_{\Phi^0{\tilde{t}}_1{\tilde{t}}_1^*}$ couplings of
eq.~(\ref{mixing}), through the $\theta_{\tilde t}$ mixing angle,
and because of the form of the 
squark-squark-Higgs vertices. 
Thirdly,
all $R(gg\to\Phi^0)$ values are
close to unity (i.e., negligible effects of the CP-violating phases) when 
$\phi_\mu$ is small for every value of $\phi_A$. This can 
easily be understood from Fig.~\ref{fig:A}, since when $\phi_\mu \ar 0$
also $|A|,\phi_{\tilde{t}} \ar 0$, so that
 no enhancement occurs in the 
$\lambda_{\Phi^0{\tilde{t}}_1{\tilde{t}}_1^*}$ couplings of
eq.~(\ref{mixing}). The same does not happen for the opposite condition
($\phi_A\to0$ for any $\phi_\mu$), since 
$|\mu|$ here is fixed and thus $\phi_{\tilde{t}}$ is always finite
when $\phi_A$ approaches zero, see Ref.~\cite{bigone}.
\begin{figure}[p]
\epsfig{figure=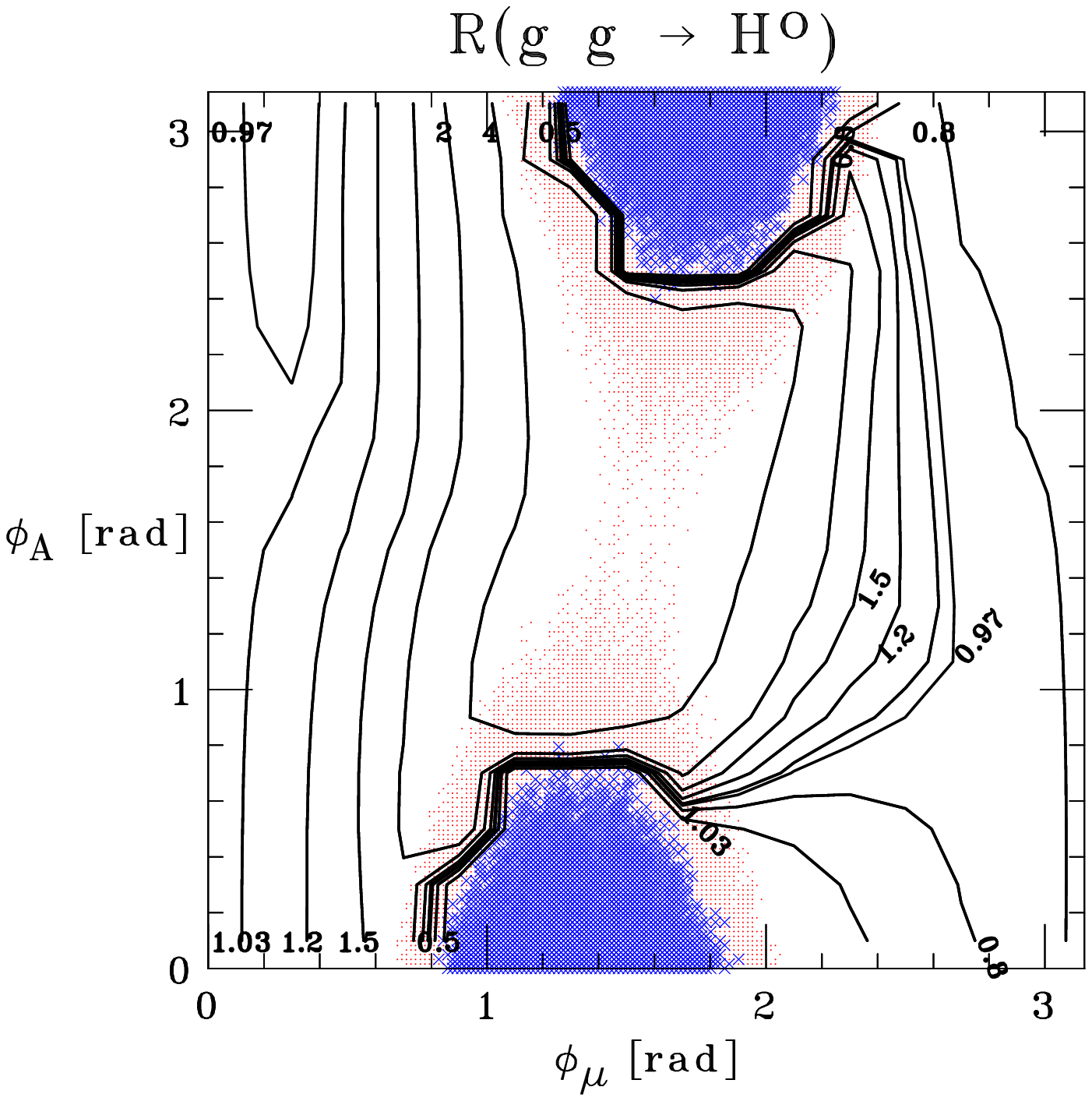,height=3.0in,angle=90}
\caption{Same as in Fig.~\ref{fig:h0} for the case $\Phi^0=H^0$.}
\label{fig:H0}
\end{figure}
\vspace{-5mm}
\begin{figure}[p]
\begin{center}
\epsfig{figure=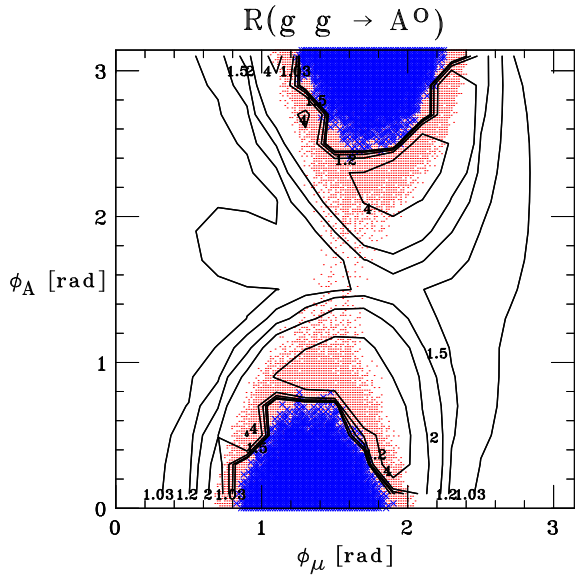,height=3.0in,angle=90}
\caption{Same as in Fig.~\ref{fig:h0} for the case $\Phi^0=A^0$.}
\label{fig:A0}
\end{center}
\end{figure}
\vspace{-3mm}
\noindent
In conclusion, we have demonstrated the potentially dramatic effects that
the presence of unconstrained (from the fermionic EDMs) CP-violating
phases in the soft SUSY sector of the MSSM can have on the dominant
-- over most of the parameter space of the model --
production mode of all neutral Higgs bosons at the LHC. 
In fact, corrections induced to the total production cross sections 
by finite values of $\phi_\mu$ and $\phi_A$ have been seen to be
much larger than any other known effect, such as higher
order EW and QCD corrections, at least for certain  combinations of
soft SUSY masses and couplings.
We feel that the matter raised here deserves further attention, both
theoretically and experimentally. To this end, a more complete analysis,
including a wider selection of combinations of MSSM parameters as 
well as the incorporation of the dominant two-loop QCD effects, is now
under completion \cite{bigone}. Similarly, one should
investigate the effect of the CP-violating phases in the decay process 
$h^0\to\gamma\gamma$ \cite{decay}, as it represents the 
most promising discovery channel of the lightest Higgs boson.

\vspace{2.0mm}
\noindent
{\small{\sl SM acknowledges
financial support from the UK PPARC and
AD from the Marie Curie Research Training Grant
ERB-FMBI-CT98-3438. The authors thank J. Rosiek for making
his code available and for his assistance in using it.}}

\end{document}